\documentclass{article}
\usepackage{graphicx} 
\usepackage{amssymb}
\usepackage[utf8]{inputenc}   
\usepackage[T1]{fontenc}      
\usepackage{amsmath}
\usepackage{amsthm}
\usepackage{color}
\usepackage{comment}
\usepackage{enumitem}		
\usepackage{geometry}		
\usepackage[percent]{overpic}
\usepackage{graphicx}
\usepackage{authblk}
\usepackage{amssymb}
\usepackage{listings}
\usepackage{xcolor}
\usepackage{caption}
\usepackage{color}
\usepackage{algorithm}
\usepackage{algpseudocode}
\usepackage{subcaption}
\usepackage{hyperref}
\usepackage{url}

\title{Effect of Levy noise on the network of coupled Chialvo neurons: Impact of topology}
\author[1]{Swetha P}
\author[1]{J. S Ram\thanks{Corresponding author (J.S. Ram): \url{ eujaidev28@gmail.com}}}
\author[1]{D.S. Varghese}
\author[1]{A.S. Nair}
\author[1]{A.M. Suresh}
\author[1]{C. Davis}
\author[1]{Remya CR}
\author[1]{Abhirami A S}
\author[1]{A. Hareendran}
\author[1]{Fathima N}
\author[1]{S.S. Muni}
\author[2]{A.V. Bukh}

\affil[1]{School of Digital Sciences, Digital University Kerala, Technopark phase-IV Campus, Mangalapuram, 695317, Thiruvananthapuram, India}
\affil[2]{Saratov State University, 83 Astrakhanskaya Street, Saratov 410012, Russia}
\date{}

\graphicspath{{figures/}}
\usepackage[english]{babel}

\begin{document}

\maketitle

\begin{abstract}
    This research investigates the complex spatiotemporal behaviors of Chialvo neuron maps under the influence of  Levy noise on three different network topologies: a ring network, a two-dimensional lattice affected by electromagnetic flux, and a delayed coupled lattice. On the ring structure, we show that adding non-uniform Levy noise induces the formation of new collective dynamics—like standing and traveling waves. The frequency and type of these emergent patterns depend sensitively on the intrinsic excitability parameter and the noise intensity, revealing new pathways to control synchronization behavior through noise modulation. In the 2D lattice network, we show that electromagnetic flux and noise together induce a diverse range of behaviors, from synchronized waves to desynchronized states. Most strikingly, spiral wave chimeras emerge under moderate noise, with coherent and incoherent regions coexisting, highlighting the fine balance between external forcing and stochastic perturbations. Finally,  upon introducing delay in the lattice structure, the system displays a rich variety of dynamical regimes—such as labyrinth patterns, rotating spirals, and target waves—whose stability and transitions are greatly affected by both delay and coupling strength.
\end{abstract}

\section*{Introduction}
\label{sec:intro}

Dynamical networks are important tools used for simulating the behavior of complex dynamical systems in varied fields, which includes neuroscience \cite{frohlich_network_2016}, power grids \cite{hill_power_2006}, cell biology \cite{thurner_anomalous_2003},  traffic flow~\cite{LU2004281, agarwal_dynamic_2016}. Dynamical networks give us the ability to study natural phenomena that cannot be understood by studying the individual entities in the network~\cite{han_dynamic_2022}. In neuroscience, networks are used to study the brain's structural and functional activity. To understand a dynamical network like the brain, it is necessary to study the structure of the connections between neurons and also the activity of individual neurons and synapses~\cite{maslennikov_adaptive_2017}.

Neurons are the basic building blocks of the nervous system which transmit information through a process known as electrochemical signalling~\cite{GORIELY201579, zhang_basic_2019}. The development of mathematical representations for neurons makes it easier to analyze the activity of the brain computationally~\cite{gerstner2002spiking}. The foundation of computational neuroscience were laid in 1952 when Hodgkin and Huxley determined the contribution of sodium and potassium ions in a neuron's axon and examined its action potential~\cite{hodgkin1952quantitative}. The following models are also widely used: Hindmarsh--Rose model~\cite{hindmarsh_model_1982}, FitzHugh--Nagumo model~\cite{fitzhugh_mathematical_1955, nagumo_active_1962}, integrate-and-fire model~\cite{burkitt_review_2006}, and Morris--Lecar model~\cite{morris1981voltage, tsumoto_bifurcations_2006, fatoyinbo_influence_2022}. To speed up calculations and simplify models researchers have proposed several discrete-time neuron models such as: Chialvo maps~\cite{CHIALVO1995461}, Rulkov maps~\cite{rulkov_modeling_2002}, discrete Izhzikevich maps~\cite{Bukh2023}, etc.  One notable study \cite{Shepelev2021} delved into synchronization phenomena within a two-layer multiplex network made up of two-dimensional lattices of van der Pol oscillators. Their findings highlighted the intricate interplay between spiral wave structures in the attractively coupled layer and labyrinth-like formations in the repulsively coupled layer. Interestingly, they observed that the emergence and stability of these patterns were highly sensitive to the strength and range of intra-layer coupling.
In another study, the focus shifted to Chialvo neuron dynamics within a minimal higher-order ring-star network influenced by diffusive couplings \cite{Nair2024}. By experimenting with the higher-order coupling strength, the authors explored collective behavior through synchronization metrics, pointing towards potential implications for understanding neural information processing and related disorders. The impact of external Levy noise on spiral wave patterns in a lattice of Chialvo neurons was explored \cite{Kolesnikov2025}. Through a thorough two-parameter analysis, they demonstrated that both the coupling structure and the characteristics of the noise played crucial roles in determining the stability and shape of spiral waves. A particularly interesting finding was the transition from spiral to target waves under certain noise conditions, revealing how noise can induce shifts and create multiple stable states in spatiotemporal dynamics. In \cite{Ramrezvila2024}, authors examined the Chialvo neuron map, mapping out parameter spaces according to periodicity and Lyapunov exponents showcasing fascinating pseudofractal structures and a rich array of periodic and chaotic attractors, including “eyes of chaos” that echo patterns found in continuous systems.

Neuronal networks with simple topologies such as the ring network~\cite{Saltzer1980, omelchenko_collective_2022}, star network~\cite{Sun2009, yang_synchronization_2022}, ring-star network~\cite{muni_discrete_2022}, two-dimensional lattice~\cite{Giordano2005, bukh_spiral_2019}, multilayer~\cite{Kivela2014}, multiplex\cite{Zhuang2017}, random~\cite{Barabasi1999}, and small-world~\cite{Zippo2013} demonstrate many rich spatiotemporal patterns, including chimera states~\cite{andreev_chimera_2019, Muni2020}, spiral waves~\cite{rothkegel_multistability_2009, feng_spiral_2019, wu_pattern_2022, souza_spiral_2024, rybalova_spiral_2019}, and target waves~\cite{qin_autapse-induced_2014, rybalova_levy_2024}. Spiral waves have been experimentally observed in the mammalian neocortex and can contribute to seizures~\cite{huang_spiral_2010, wu_propagating_2008}. Apart from neuronal networks, spiral waves have also been observed in various other biological systems. Spiral wave patterns in heart tissue have been linked to heart conditions like ventricular fibrillation~\cite{samie_mechanisms_2001}. Target waves have been found to disperse spiral waves and reduce the effect of ion channel block in neuronal networks~\cite{ma_emergence_2013}. The spatiotemporal patterns, as a neural activity organization in time and space, become relevant, meaningful patterns, allowing the brain to render complex information such as the direction of movement or visual scenes changing where all responses from neurons will be altered according to the input's spatial and temporal change~\cite{Kurogi1987}.

Excitability is the ability of neurons to respond to stimuli~\cite{rutecki_neuronal_1992}. The effect of external stimuli such as electromagnetic radiation, chemicals, optical signals, etc. on neuronal activity has been studied~\cite{feng_route_2017, yan_further_2020, hussain_synchronization_2021, muni_discrete_2022, FRANASZCZUK200365}. The nervous system is often subjected to external disturbances. Effect of external stimuli like noise~\cite{Vinaya2018, Bashkirtseva2023}, electromagnetic flux~\cite{muni_dynamical_2022}, strong pulse~\cite{Ram2024} on network of neurons were extensively studying recently. Levy noise~\cite{Patel2008, Liu2019} and Gaussian noise~\cite{Gao2021}, particularly, are two major stochastic processes that have unique features leading to different effects on neuronal functioning. Levy noise with stability parameter $\alpha<2$ is a non-Gaussian, heavy-tailed noise with stochastic jumps~\cite{wu_levy_2017}. Unlike Gaussian noise, which has a finite variance and can only generate small fluctuations around an average value, Levy noise can generate giant, random perturbations. It can thus properly capture the irregularities and sudden changes that may appear in complex biological systems~\cite{wu_levy_2017, Li2022}. Levy noise is able to model such an unpredictability and sudden change caused by stochastic leaps~\cite{Li2022}, which is really useful in researching neurological phenomena like a chimera state, when coherent and incoherent behaviours coexist. In particular, Levy noise reproduces the noise in real-world neurons by introducing frequent but essential fluctuations that enhance weak signals through stochastic resonance and broaden the range of firing patterns within the neurons~\cite{Patel2008}.

Time delay, one such external influence, is an intrinsic property in neuronal networks. The finite propagation velocity of action potential along the axons, as well as time lapses in signal transmission (synaptic process) and reception (dendritic process) between neurons, create time delays in the network of neurons~\cite{MadadiAsl2018}. Time delays also play an important role in various fields. Several important studies have laid the groundwork for understanding how time delays affect neural and oscillatory systems. For instance, how timelike delays in neural networks can result in chaotic and synchronization behaviors which are crucial for information processing in the brain, is discussed in~\cite{Zhou2002}. In~\cite{Masoller2005}, the influence of random delays on synchronizing an array of coupled chaotic logistic maps are studied. The study reveals that, for adequate coupling strength, the array can synchronize in a steady state, and the chaotic dynamics of individual maps are suppressed. The role of time delays in the stability of coupled discrete-time systems~\cite{Atay2011}, discrete-time recurrent neural networks~\cite{YuZhao2009}, a class of neural networks with Lipschitz activation function~\cite{Solak2023} has been reported in the literature. How time delay in coupled neuron models serves as a mechanism to suppress or promote synchronization depending on the delay duration was highlighted in~\cite{Hansen2022, Schll2009}. Many recent research studies have been devoted to delay-induced phenomena~\cite{Popovych2011, Balanov2004, Balanov2006}. It is well known that delay can lead to the destabilization of a stationary point and result in oscillatory behavior~\cite{DAHLEM2009}. Additionally, it has been demonstrated that the dimension of chaotic attractors in delay systems is directly proportional to the magnitude of delay \cite{DoyneFarmer1982}. This indicates that delay differential equations with high delays can exhibit high dimensional chaos, making them useful for various applications.

In this research, we have considered three different network topologies of Chialvo neuron mappings under the consideration of L\'{e}vy noise. In particular, we have considered:(a) the response of Chialvo neurons to Levy noise in a ring network topology, (b) the effect of electromagnetic flux and Levy noise on the behavior of a two-dimensional array of Chialvo neurons, and (c) the effect of delay on a lattice network of Chialvo neurons. The main novel contributions of the research are as follows:
\begin{itemize}
\item Response of Chialvo neurons to L\'{e}vy noise in ring network topology
\begin{itemize}
    \item Introduction of non-uniform L\'{e}vy noise fundamentally alters the spatiotemporal dynamics of the Chialvo neuron ring, inducing novel patterns such as standing and traveling waves, which are absent under noise-free conditions.
    \item Noise significantly affects the mean oscillation frequency in a specific range of the intrinsic parameter $b$, especially between 0.18 and 0.22, thereby reshaping the regime landscape to include incoherence, regular oscillations, and wave-like dynamics.
    \item Both the spatial structure and intensity of noise can serve as control parameters to shift the dynamical regime in the $(b,\gamma)$ parameter space, enabling modulation of neuronal behavior in the ring topology.
\end{itemize}
\item Effect of electromagnetic flux and L\'{e}vy noise on 2D lattice of Chialvo neuron map
\begin{itemize}
    \item The analysis of the regime map contained some target wave pattern at lower values of flux, while at higher flux values, synchronized states were dominant with a few regions displaying unsynchronized states.
    \item  Levy noise was shown to interfere with the normal states and thus the system behaved with a mixture of synchronized and desynchronized states with this feature becoming more evident with increasing levels of noise.
    \item  Of particular interest is that at moderate levels of noise, states of spiral wave chimeras were formed, with portions of the system propagating a wave and other portions being completely asynchronized.
    \item Noise levels beyond this worsened the dispersion of the dynamics but a spiral wave and many centered spiral states survived indicating how much external electromagnetic interference as well as noise affects neural activities.
\end{itemize}

\item Effect of delay on a lattice network of Chialvo neurons
\begin{itemize}
    \item  The study reveals that varying coupling strength $\gamma$ and time delay $\tau$ in a lattice of Chialvo neuron maps gives rise to a variety of dynamical regimes, including synchronized states, target waves, incoherence, labyrinth-like patterns, and spiral waves.

\item Time delays significantly impact the transitions between dynamical regimes. For instance, labyrinth-like structures and spiral waves appear and disappear periodically with changes in delay, emphasizing its critical role in shaping spatiotemporal patterns.

\item Specific regimes, like multi-center target waves and smooth profiles, are particularly sensitive to variations in delay and coupling strength. These patterns demonstrate complex interference effects and highlight the role of phase mismatches introduced by delays.

\item The prevalence of incoherent regimes at moderate coupling and higher delays suggests that delay-induced desynchronization could have functional implications, offering insights into both biological and artificial neural networks.
\end{itemize}
\end{itemize} 

In Section \S \ref{sec:model}, we introduce the three network topologies used for implementing the Chialvo map: (a) a ring network of coupled Chialvo neurons (\S \ref{sec:RingIntro}), and (b) a two-dimensional lattice of coupled Chialvo neurons (\S \ref{sec:2DChialvo}). \S \ref{sec:res} presents the main results for each topology: (a) spatiotemporal patterns in the ring network (\S \ref{sec:ringres}), (b) the effect of electromagnetic flux on the 2D lattice (\S \ref{sec:emfluxres}), and (c) the impact of time delay on the spatiotemporal dynamics of the 2D lattice (\S \ref{sec:delay2D}). The paper concludes with a summary of findings and potential directions for future work.
\section{Model under study and initial regime}\label{sec:model}
A single Chialvo neuron with electromagnetic flux is described by the following equations:
\begin{equation}\label{singleChialvo}
\begin{aligned}
    x(n+1) &= f(x(n), y(n), \phi(n)) = (x(n))^2 e^{y(n) - x(n)} + k_0 + k x(n) M(\phi(n)), \\
    y(n+1) &= g(x(n), y(n), \phi(n)) = a y(n) - b x(n) + c, \\
    \phi(n+1) &= q(x(n), y(n), \phi(n)) = k_1 x(n) - k_2 \phi(n),
\end{aligned}
\end{equation}
where $x$ and $y$ are the activation potential variable and recovery potential variables respectively, and $\phi$ represents the magnetic flux across the neuron membrane. The model has the several dynamical parameters: $k_0$ can be a constant bias term or a time-dependent additive perturbation, $a<1$ is the time constant of recovery, $b<1$ characterizes the activation-dependence of the recovery process, $c$ represents an offset constant. The term $kxM(\phi)$ represents the induction current resulting from changes in the magnetic flux~\cite{muni_dynamical_2022, wu_levy_2017}. The parameter $k$ is referred to as the feedback gain, and $M(\phi)$ denotes the memconductance of the memristor controlled by the magnetic flux. In this study, we adopt a commonly used memconductance function given by $M(\phi) = \alpha + 3\beta\phi^2$ with $a = 0.1$ and $b = 0.2$.

\subsection{Ring of coupled Chialvo neurons}
\label{sec:RingIntro}
The dynamical equation for the ring of coupled Chialvo neurons is presented as follows:
\begin{equation}\label{ringChialvo}
\begin{aligned}
    x_m(n+1) &= f(x_m(n), y_m(n), \phi_m(n)) + \eta_m^{\sigma,\alpha,\beta}(n) + \\
             &+ \frac{\gamma}{2R} \sum_{i=m-R}^{m+R} \left( x_i(n-\tau) - x_m(n) \right), \\
    y_m(n+1) &= g(x_m(n), y_m(n), \phi_m(n)), \\
    \phi_m(n+1) &= q(x_m(n), y_m(n), \phi_m(n)),
\end{aligned}
\end{equation}
where functions $f, g$, and $q$ are defined in~\eqref{singleChialvo}, $m$ is a node index, the boundary conditions are periodic ($i \pm N=i$). The neighborhood coupling of the system is being explained by the last term in the first equation in~\eqref{ringChialvo}. The interaction range around the $m$th node is represented by the parameter $R$. The strength of coupling within the neighborhood is represented by $\gamma$. The delay in coupling is expressed with $\tau$. The Levy noise that influences the neurons is represented by $\eta_m^{\sigma,\alpha,\beta}(n)$. Here, $\sigma$, $\alpha$, and $\beta$ characterize the scale, stability, and asymmetry of the Levy noise sources. For all simulations of the ring network, the initial conditions are set to the fixed point state for each Chialvo neuron. The introduction of external noise induces spiking activity across the network.

\subsection{2D lattice of coupled Chialvo neurons}
\label{sec:2DChialvo}
\begin{figure}[!ht]
\centering
\includegraphics[width=0.85\linewidth]{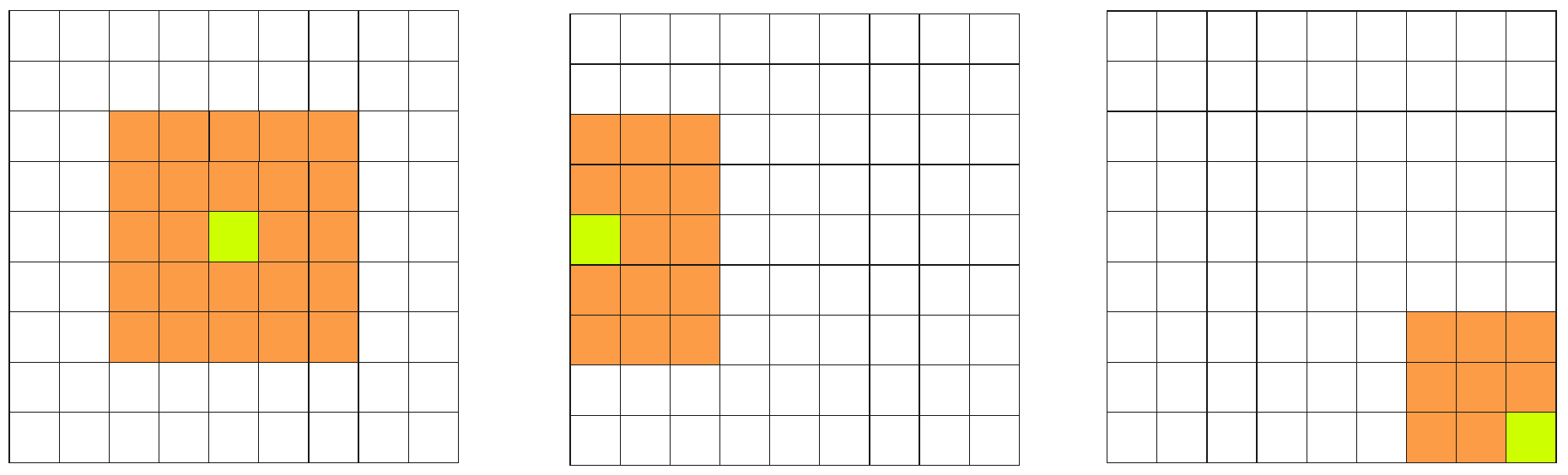}\\
(a)~~~~~~~~~~~~~~~~~~~~~~~~~~~~~~~~~~ (b) ~~~~~~~~~~~~~~~~~~~~~~~~~~~~~~~~~~(c)
\caption{Schematic representation of the nonlocal coupling with no-flux boundary conditions defined in~\eqref{eq:Noflux}. Schemes of the coupling in~\eqref{eq:LatticeChialvo} for different locations of the selected node (marked in green): (a) in the lattice center, (b) in the middle of the left edge, and (d) in the right bottom corner. Oscillators coupled with a selected green ($i, j$)th node are marked in orange, and the remaining uncoupled nodes are shown in white.}
\label{fig:no-flux}
\end{figure}

The dynamical equation for the 2D lattice of coupled Chialvo neurons is expressed with the following relations:
\begin{equation}   
\begin{aligned}
   x_{i, j}(n+1) &=f(x_{i, j}(n), y_{i, j}(n), \phi_{i, j}(n)) + \eta_{i, j}^{\sigma,\alpha,\beta}(n) + \\
   &+ \frac{\gamma}{B_{i, j}} \sum_{m, n \in Q_{i, j}}\left(x_{m, n}(n - \tau)-x_{i, j}\right), \\
   y_{i, j}(n+1) &=g(x_{i, j}(n), y_{i, j}(n), \phi_{i, j}(n)), \\
   \phi_{i, j}(n+1) &= q(x_{i, j}(n), y_{i, j}(n), \phi_{i, j}(n)),
\end{aligned}
\label{eq:LatticeChialvo}
\end{equation}
where functions $f, g$, and $q$ are defined in~\eqref{singleChialvo}, $i$ and $j$ are the node indices on the lattice, the boundary conditions are no-flux with the following $(m,n)\in Q_{i,j}$:
\begin{equation}
\begin{aligned}
    m = \max(i-P,0),\dots,\min(i+P,N),\\
    n = \max(j-P,0),\dots,\min(j+P,N),
\end{aligned}
\label{eq:Noflux}
\end{equation}
with coupling range $P$. We are considering an $N\times N$ lattice, where $N=50$. An illustration of the no-flux boundary conditions used in this study is shown in Fig. \ref{fig:no-flux} for various locations of the $i,j$th oscillator for the coupling range $P=2$.

\begin{figure}[!ht]
    \centering
    \includegraphics[width=0.5\textwidth]{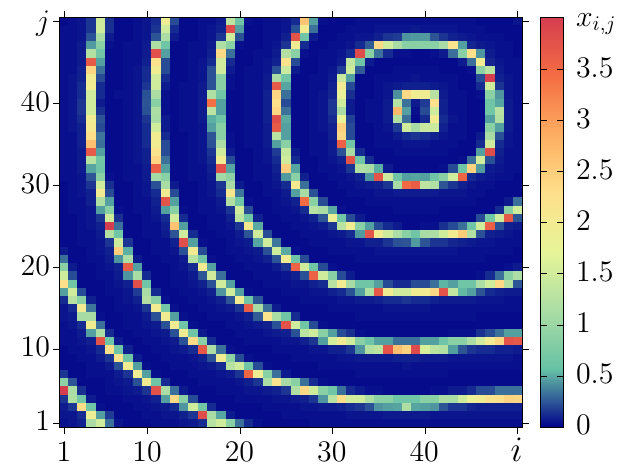}
    \caption{The figure shows a target wave in the 2D lattice \eqref{eq:LatticeChialvo}. This has been used as the initial condition for numerical simulation results. Parameters are fixed as: $a = 0.89$, $c = 0.26$, $k_0 = 0.031$, $\gamma = 0.0015$, $\alpha_L=2$, $\beta_L=0$, $k_1=0.1$, and $k_2=0.2$.} 
    \label{fig:IC}
\end{figure}

In this paper, for the lattice model~\eqref{eq:LatticeChialvo} we use the following values for the dynamical parameters to get the initial spatio-temporal structure representing a target wave: $a = 0.89$, $c = 0.26$, $k_0 = 0.031$, $\gamma = 0.0015$, $\alpha_L=2$, $\beta_L=0$, $k_1=0.1$, and $k_2=0.2$. The target wave is shown in Fig.~\ref{fig:IC}. Our goal is to study the effects of the noise and delay influence on this structure.

\section{Results}
\label{sec:res}
We have organized the main results on the spatiotemporal patterns arising in three different network topologies of Chialvo neuron mapping. In Section \ref{sec:ringres}, we examine a ring network of Chialvo neurons and highlight the similarities between the regime maps and frequency plots under varying noise levels, demonstrating transitions between distinct spatiotemporal regimes. Section \ref{sec:emfluxres} focuses on a 2D lattice of Chialvo neurons influenced by an external electromagnetic flux, where we present the main regime map and illustrate the range of spatiotemporal patterns observed. In Section \ref{sec:delay2D}, we analyze the impact of time delay on the spatiotemporal dynamics in the same 2D lattice setup and showcase almost periodic structures in the regime map.
\subsection{Spatiotemporal Dynamics in a Ring Network of Chialvo Neurons}
\label{sec:ringres}
The ring network of coupled Chialvo neurons introduced in \S \ref{sec:RingIntro} displays a plethora of spatiotemporal patterns with the presence of noise and time delay. Here we analyze the dynamical effects of the presence of noise and time delay on the transition between various spatiotemporal patterns. 
\begin{figure}[!htbp]
    \centering
    \begin{tabular}{ccc}
        \begin{overpic}[width=0.3\linewidth]{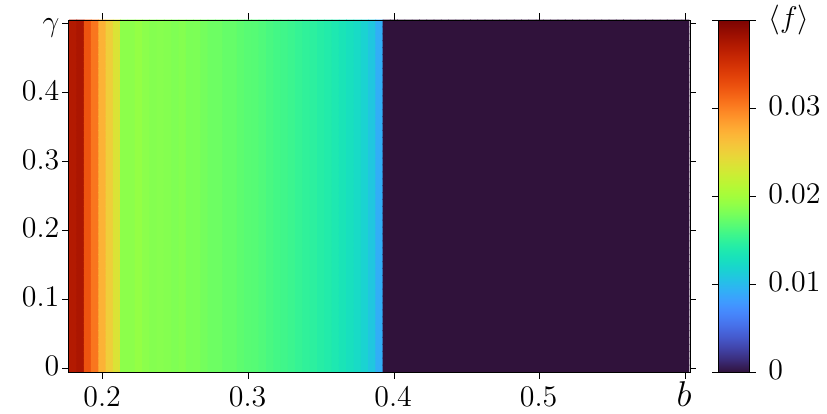}
        \put(-10,50){\text{(a)}}
        \end{overpic} &
        \begin{overpic}[width=0.3\linewidth]{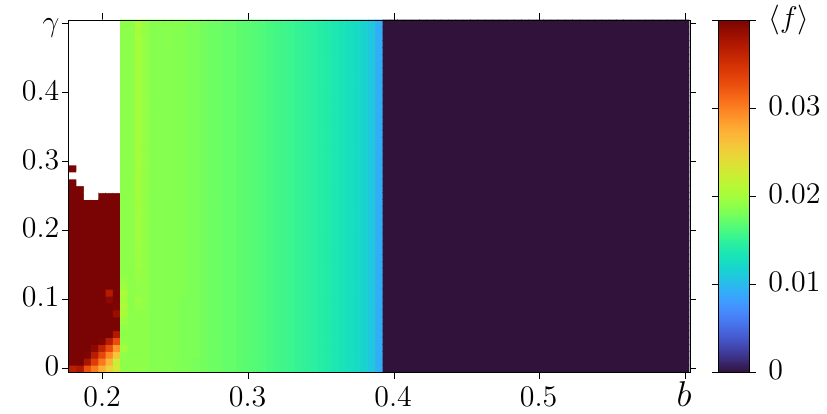}
         \put(-10,50){\text{(b)}}
        \end{overpic} &
        \begin{overpic}[width=0.3\linewidth]{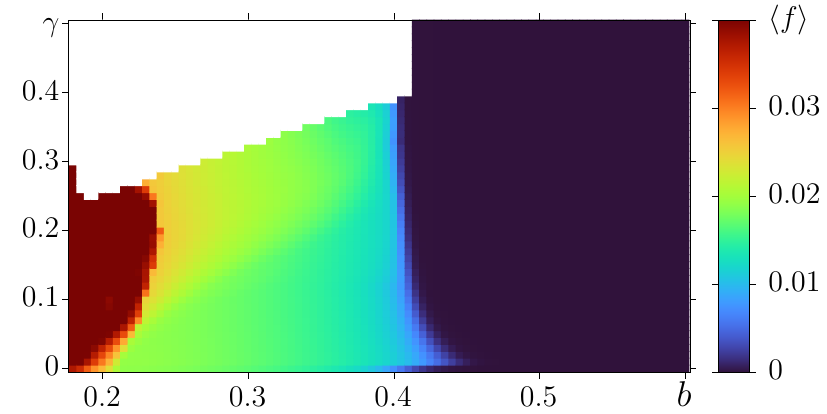}
         \put(-10,50){\text{(c)}}
        \end{overpic}\\
        \begin{overpic}[width=0.3\linewidth]{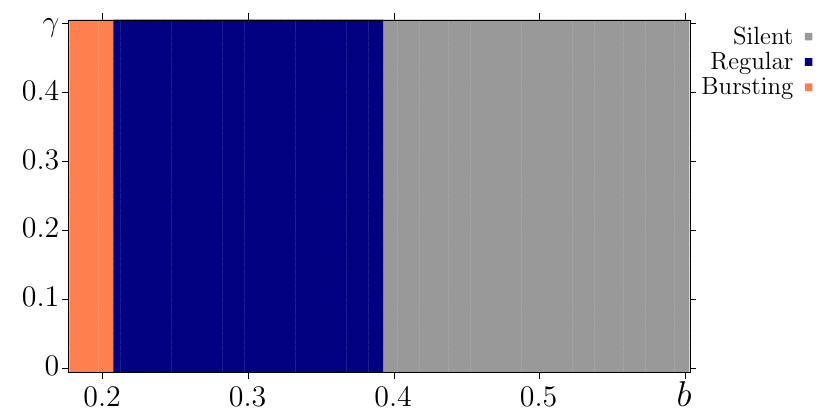}
         \put(-10,50){\text{(d)}}
        \end{overpic} &
        \begin{overpic}[width=0.3\linewidth]{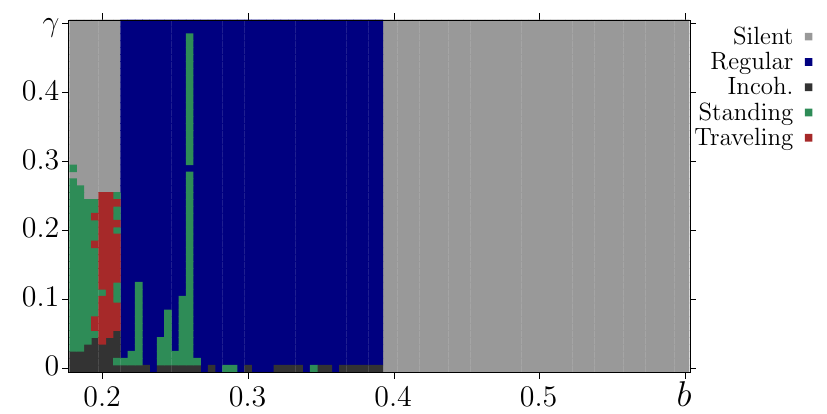}
         \put(-10,50){\text{(e)}}
        \end{overpic} &
        \begin{overpic}[width=0.3\linewidth]{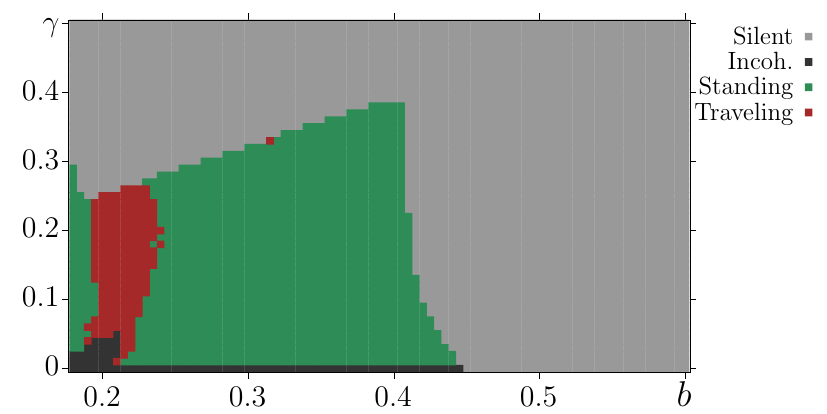}
         \put(-10,50){\text{(f)}}
        \end{overpic}
    \end{tabular}
    \caption{The frequency maps (a-c) and the regime maps (d-f) on the $(b, \gamma)$ parameter plane for different values of noise scale parameter $\sigma$=0 (a, d), $\sigma=10^{-4}$ (b, e), $\sigma=10^{-2}$ (c, f) with $a=0.89, c=0.28, k=0.025,$ and $P=1$.}
    \label{fig:RingRegimeFreq}
\end{figure}

\begin{figure}[!htbp]
    \centering
    \begin{tabular}{ccc}
        \begin{overpic}[width=0.3\linewidth]{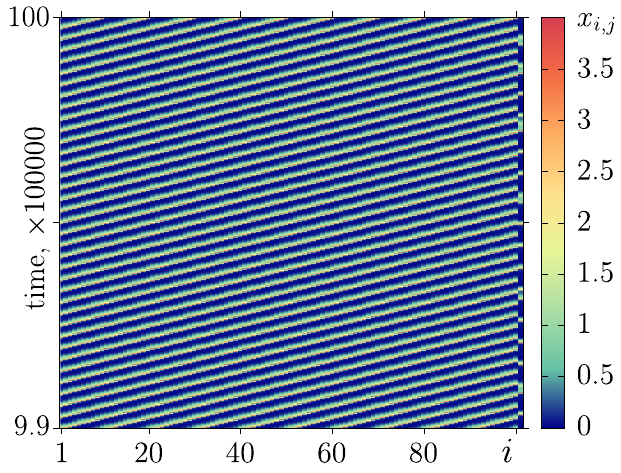}
        \put(-10,85){\text{(a)}}
        \end{overpic} &
        \begin{overpic}[width=0.3\linewidth]{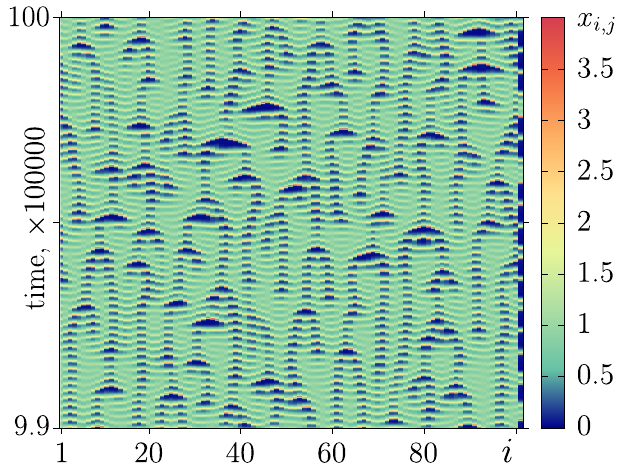}
        \put(-10,85){\text{(b)}}
        \end{overpic} &
        \begin{overpic}[width=0.3\linewidth]{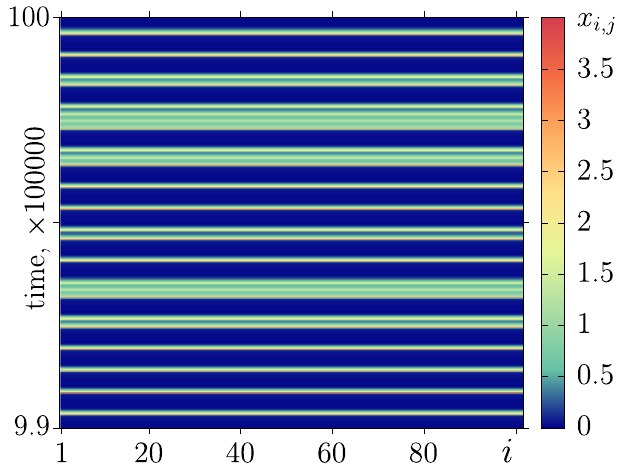}
       \put(-10,85)
       {\text{(c)}}
        \end{overpic}
    \end{tabular}
    \caption{ Space-time diagrams for travelling wave regime at $b=0.195, \gamma=0.18, \sigma=10^{-5}$ (a), standing wave regime at $b=0.18, g=0.15, \sigma=10^{-5}$ (b), and bursting regime at $b=0.18, g=0.09, \sigma=0$ (c) observed in the ring of Chialvo neurons with $a=0.89, c=0.28, k=0.025,$ and $P=1$.}
    \label{fig:ringpatterns}
\end{figure}
The mean oscillation frequency maps for the Chialvo neuron ring shown in Fig. \ref{fig:RingRegimeFreq} reveal that in the absence of noise (Fig. \ref{fig:RingRegimeFreq} a) the coupling strength between the Chialvo neurons in the ring does not affect their spiking activity. The behavior of the Chialvo neurons in this case is controlled only by the parameter b. At the same time, the analysis of the emerging regimes shows that the entire considered range of parameter values is divided into three, including the regimes of no oscillations (Silent), periodic oscillations (Regular), and bursting oscillations (Bursting) (Fig. \ref{fig:RingRegimeFreq} d), see Fig. \ref{fig:ringpatterns} (c) where an example of the space-time diagrams is shown for the bursting regime. Introducing non-uniform noise into the Chialvo neurons coupled in the ring leads to a change in the mean oscillation frequency map in the range of parameter $b$ values from 0.18 to 0.22. The oscillation frequency increases significantly at coupling strength parameter values less than 0.25, and at coupling strength parameter values greater than 0.25, the noise leads to the destruction of the neuron model (phase trajectory goes to infinity). At the same time, the rest of the mean frequency map remains unchanged (Fig. \ref{fig:RingRegimeFreq} b). The regime map undergoes significant changes not only in the range of b values from 0.18 to 0.22, but also at b values greater than 0.22, noise-induced standing waves and incoherent oscillations arise in the ring of Chialvo neurons, which were not detected in the absence of noise. In addition, areas of standing waves, traveling waves and incoherent modes arise in the range of b values from 0.18 to 0.22 as a result of the noise effect on the Chialvo neurons coupled in the ring (Fig. \ref{fig:RingRegimeFreq} e). Space-time diagrams of traveling and standing waves are shown in Fig. \ref{fig:ringpatterns} (b) and (c) respectively. A further increase in the noise intensity leads to a change in the average frequency map in larger ranges of parameter $b$ values (Fig. \ref{fig:RingRegimeFreq} c). In this case, the area of regular synchronous oscillations completely disappears, and the areas of standing and traveling waves increase significantly. Thus, the effect of noise on the ring of Chialvo neurons can lead to the emergence of standing and traveling wave modes. In this case, the regions of occurrence of both induced modes are shifted on the plane of parameters $(b,\gamma)$, which makes it possible to control the behavior regime of the Chialvo neuron ring using the parameters of an external noise source.

\subsection{Spatiotemporal Dynamics of Chialvo Neurons on a 2D Lattice under Electromagnetic Flux}
\label{sec:emfluxres}
With the simultaneous variation of electromagnetic flux feedback gain parameter ($k$) and the Levy noise distribution width ($\sigma^L$), a variety of diverse spatiotemporal patterns is observed below:
\label{RM_flux_sigma}
\begin{figure}[H]    
    \centering
    \includegraphics[width=1.0\textwidth]{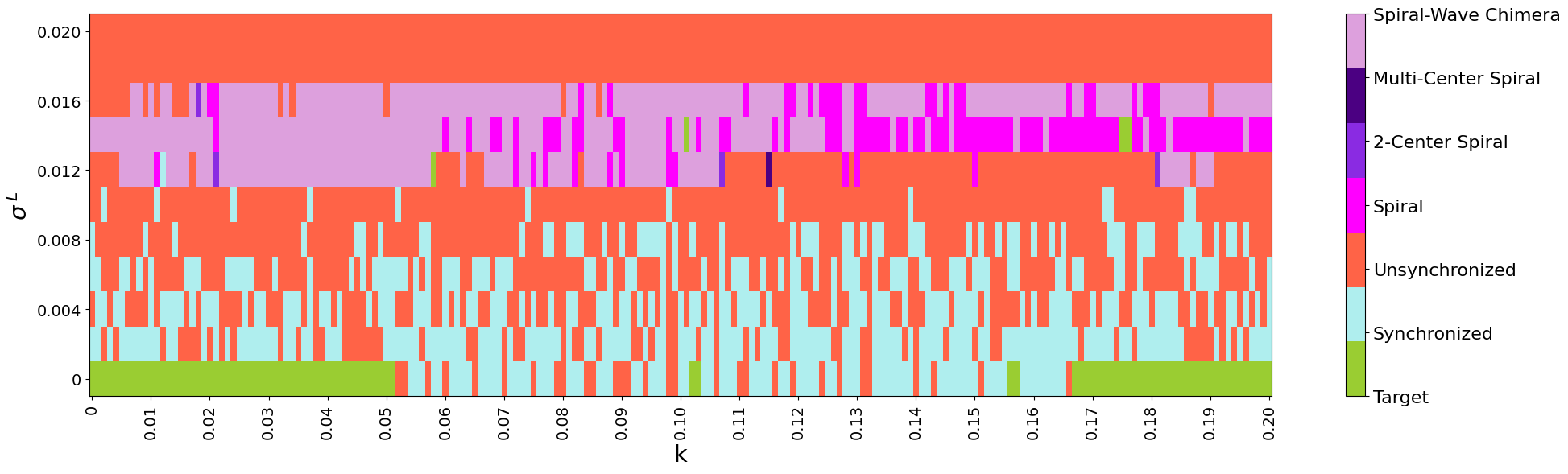}  
    \caption{Regime map plot of electromagnetic flux gain parameter $k$ vs the L\'{e}vy noise distribution width $\sigma^{L}$ color coded according to various spatiotemporal patterns shown in the legend. The plot shows the different spatio-temporal patterns observed in the 2D lattice when varying the flux parameter between 0 and 0.2 in steps of 0.001 and the sigma parameter between 0 and 0.02 in steps of 0.002.} 
    \label{fig:rm}
\end{figure}

To explore the effect of noise on the transition between various spatiotemporal patterns, we first fixed $\sigma^L = 0$ and varied the flux parameter $k$. For small $k$ values (up to 0.051), the system predominantly exhibits target wave patterns. Beyond this threshold, synchronized states with intermittent unsynchronized regions emerge. When $k$ exceeds 0.166, target waves reappear in the network lattice. For lower magnitude of noise $\sigma^L = 0.0002$, we observe that target waves vanish, even at low $k$. This indicates that the target wave regime is sensitive to noise. Instead, the system shows a mixed regime of synchronized and unsynchronized states over various $k$ values. This indicates that a  small parameter variation can lead the network system settle to synchronized or unsynchronized state indicating complexity of the network dynamics.

With further increase in strength of noise, unsynchronized states become more prevalent. For instance, at $\sigma^L = 0.0010$, unsynchronized dynamics dominate, though some synchronized patches persist. Interestingly, at $\sigma^L = 0.0012$, spiral wave chimera states emerge for low $k$, accompanied by a blend of spiral and unsynchronized states at higher $k$. For $\sigma^L = 0.0014$ and $0.0016$, we observe diverse spatiotemporal structures, including a) Two-center spirals, b) Multi-center spiral states, and c) Target waves. Notably, at $\sigma^L = 0.0018$ and $0.0020$, only unsynchronized dynamics are observed across the entire flux parameter range, suggesting a transition to full spatiotemporal disorder under strong noise.

\subsubsection{Simultaneous variation of electromagnetic flux feedback gain parameter ($k$) and the activation dependence of recovery potential parameter ($b$)}
\label{rm_flux_b}
\begin{figure}[H]    
    \centering
    \includegraphics[width=1.0\textwidth]{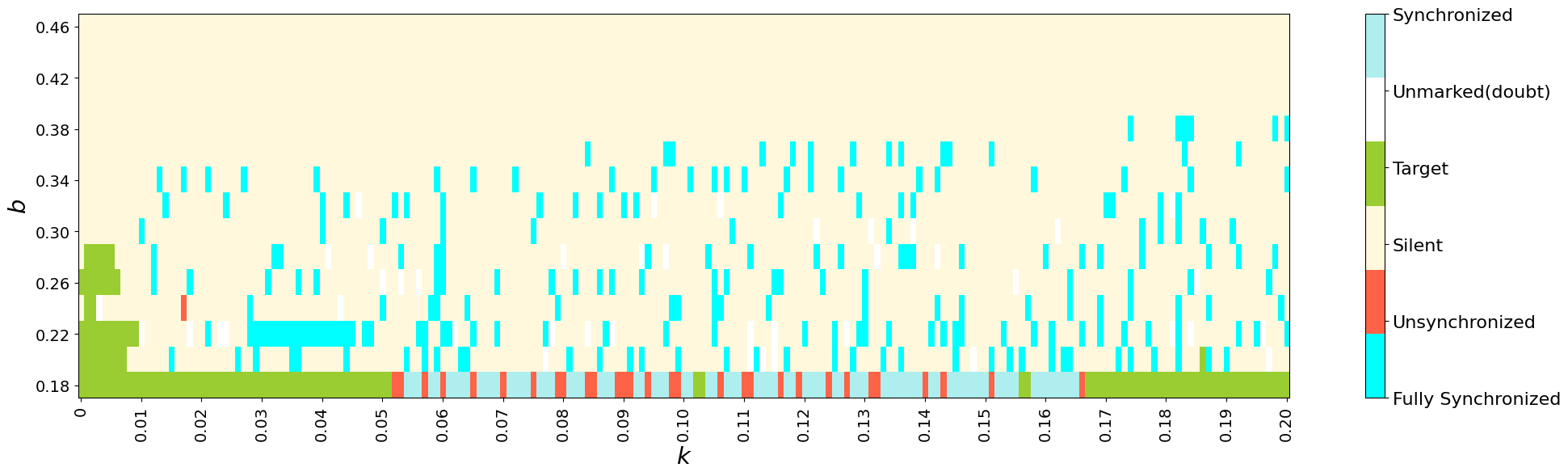}  
    \caption{Regime map plot of electromagnetic flux parameter $k$ between 0 and 0.2 in steps of 0.001 and the recovery potential variable parameter $b$ between 0.18 and 0.46 in steps of 0.02.} 
    \label{fig:rmb}
\end{figure}

The regime map demonstrates the effects of simultaneously varying the flux feedback gain parameter $k$ and the recovery potential parameter $b$ on various transitions between spatiotemporal patterns, with a focus on the resulting synchronization states. The analysis reveals distinct regions where specific dynamical states dominate. In particular, strong synchronization is observed in regions corresponding to higher values of recovery potential levels $b$, as indicated by the prevalence of fully synchronized states (cyan). The map also captures transitions between synchronized, unsynchronized, target, and silent regimes, offering deeper understanding into how the interplay between $k$ and $b$ shapes the collective behavior of the network. The distribution of the synchronized state (light blue) varies erratically across a wide parameter space, hinting at only partial coherence within the network. On the other hand, the unsynchronized state (red) appears more localized, mainly clustered in areas where both the recovery potential and flux parameters are low—suggesting a tendency toward unstable behavior in those zones. Meanwhile, target wave patterns (green) and silent states (yellow) are scarcely observed and are limited to specific regions, indicating that they arise only under particular conditions. In the following section, we explore in greater detail the range of spatiotemporal patterns produced by the 2D lattice of Chialvo neurons when exposed to Levy noise and electromagnetic flux.
\subsubsection{Spatiotemporal Patterns of the 2D lattice of Chialvo neurons}
\begin{figure}[H]
\captionsetup[subfigure]{justification=centering}
    \centering
    \begin{subfigure}[b]{0.45\textwidth}
        \includegraphics[width=\textwidth, height = 5cm, scale=1.1]{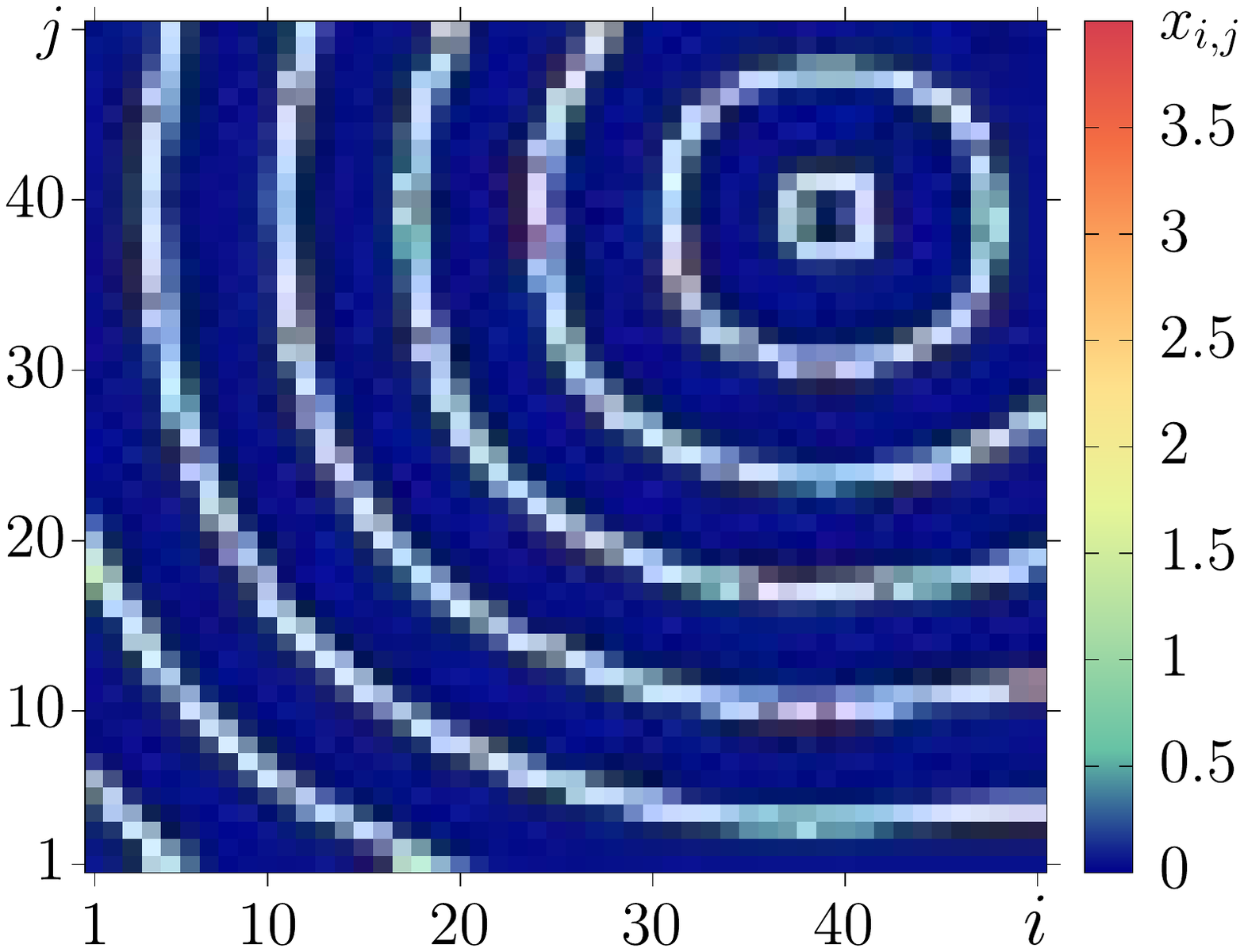}
        \caption{}
        \label{fig:targetwave_pattern}
    \end{subfigure}
    ~
    \begin{subfigure}[b]{0.45\textwidth}
        \includegraphics[width=\textwidth, height = 5cm, scale=1.1]{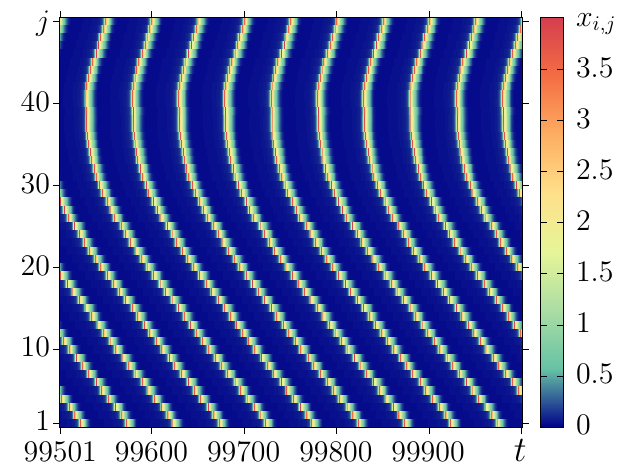}
        \caption{}
        \label{fig:target_hm}
    \end{subfigure} 
    \caption{(a) The plot shows the emergence of target waves. At $\sigma = 0.012$, a target wave can be found out and at $\sigma = 0.014$, a single target wave appears. In addition to this, between a spiral wave chimera and a spiral wave, two target waves are observed. (b) The figure represented at $i = 25$ shows a heatmap depicting the evolution of a target wave composed of concentric ring structures or wavefronts progressing outwards and sustained rhythmic oscillations in $x_{i,j}$, across the index $j$ in space and time denoted by $t$. This can be attributed to the symmetry and continuous active perturbation present at the center of the system.}
    \label{fig:target}
\end{figure}

\begin{figure}[H]
\captionsetup[subfigure]{justification=centering}
    \centering
    \begin{subfigure}[b]{0.45\textwidth}
        \includegraphics[width=\textwidth,  height = 5cm,scale=1.1]{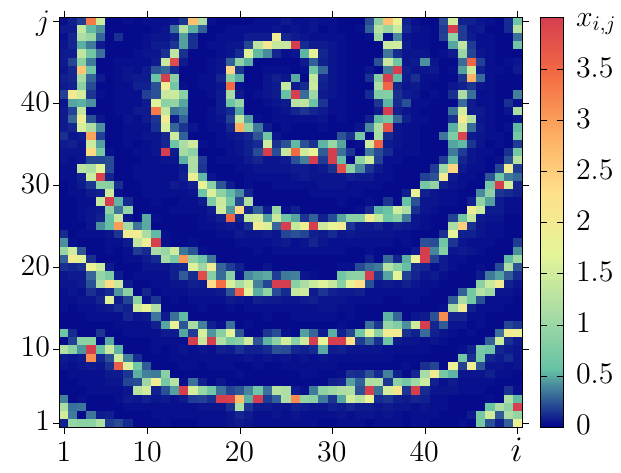}
        \caption{}
        \label{fig:spiral_pattern}
    \end{subfigure}
    ~
    \begin{subfigure}[b]{0.45\textwidth}
        \includegraphics[width=\textwidth, height = 5cm,scale=1.1]{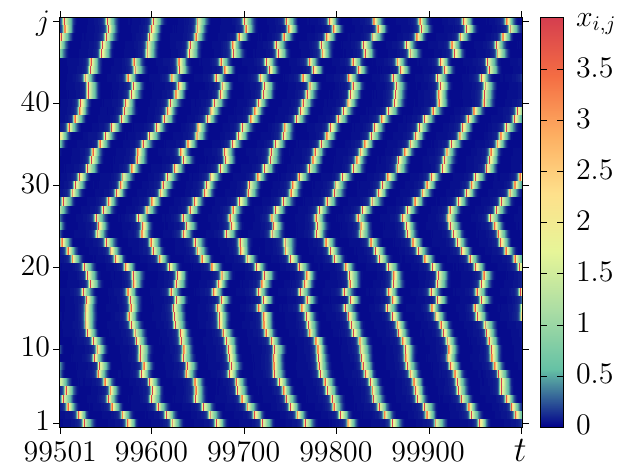}
        \caption{}
        \label{fig:spiral_hm}
    \end{subfigure} 
    \caption{(a) A spiral wave pattern emerges at $\sigma=0.0012$ and $k$ = 0.0110, highlighting the development of organized wavefronts in the lattice. The system exhibits partial synchronization, with coherent wave structures dominating the dynamics. (b) The figure depicting the heatmap captures the time dynamics of a spiral wave, exhibiting the state variable at a fixed spatial index $i = 25$. The x-axis represents time, with the y-axis corresponding to the spatial index (j), illustrating periodic band-like structures that emphasize oscillatory behavior typical for the spiral wave over time.}
    \label{fig:spiral}
\end{figure}

\begin{figure}[H]
\captionsetup[subfigure]{justification=centering}
    \centering
    \begin{subfigure}[b]{0.45\textwidth}
        \includegraphics[width=\textwidth,  scale=1.1]{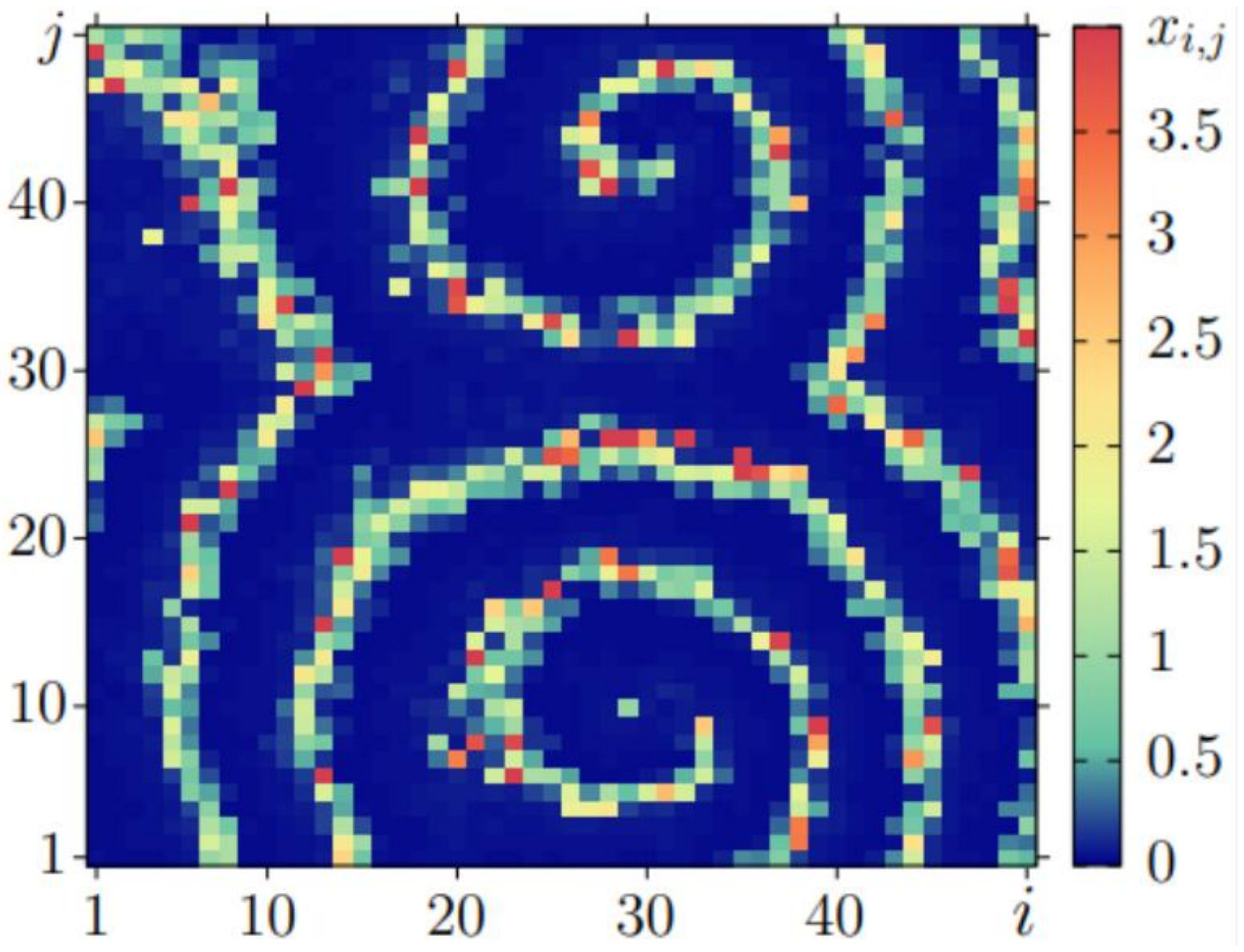}
        \caption{}
        \label{fig:2-center_pattern}
    \end{subfigure}
    ~
    \begin{subfigure}[b]{0.45\textwidth}
        \includegraphics[width=\textwidth,height = 5cm, scale=1.1]{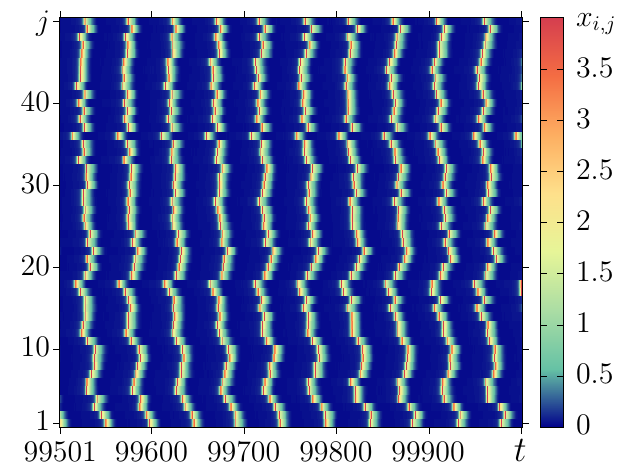}
        \caption{}
        \label{fig:2-center_hm}
    \end{subfigure} 
    \caption{(a) Two-centered spiral wave pattern appears at $\sigma = 0.0012$ and $k$ = 0.1810, with two distinct spiral cores driving the wave propagation. The elevated flux facilitates the formation of dual spirals, leading to synchronized wavefronts from both centers. (b) Figure  shows the time dynamics of a two-centered spiral wave for a fixed spatial index  $i = 25$  where the x-axis is time $t$  and the y-axis is the spatial index  $j$. The periodic band-like structures in the heatmap reflect oscillations in the wave, highlighting the matching and coherence of the two spiral centers at different instances in time. Thus, the visualization depicts the complex time organization implicit in the system. }
    \label{fig:2-center}
\end{figure}

\begin{figure}[H]
\captionsetup[subfigure]{justification=centering}
    \centering
    \begin{subfigure}[b]{0.45\textwidth}
        \includegraphics[width=\textwidth, height = 5cm, scale=1.1]{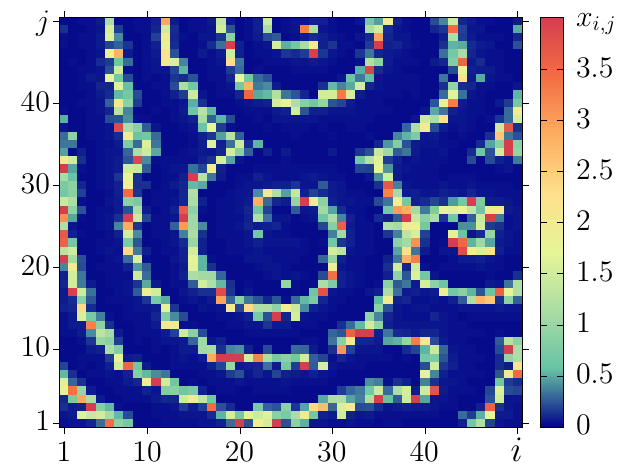}
        \caption{}
        \label{fig:multi-center_pattern}
    \end{subfigure}
    ~
    \begin{subfigure}[b]{0.45\textwidth}
        \includegraphics[width=\textwidth, height = 5cm,scale=1.1]{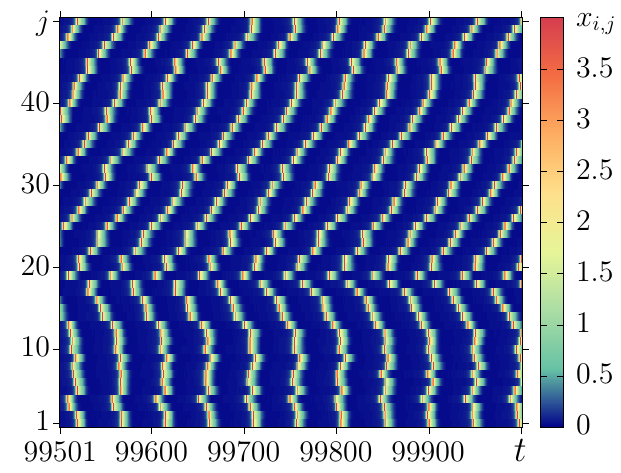}
        \caption{}
        \label{fig:multi-center_hm}
    \end{subfigure} 
    \caption{(a) multi-centered spiral wave pattern forms at $\sigma = 0.0012$ and $k$ = 0.1150, displaying several spiral cores across the lattice. The high flux value leads to multiple excitation points, creating complex, overlapping wave dynamics (b) The spatiotemporal history of multi-centered spirals is visualized as a heatmap at $i=25$, emphasizing the periodic modulation of $x_{i,j}$, with spatial index $j$ and time $t$, as being pushed along by nonlinear coupling processes coupled with an inherent feedback of dynamics.}
    \label{fig:multi-center}
\end{figure}

\begin{figure}[H]
\captionsetup[subfigure]{justification=centering}
    \centering
    \begin{subfigure}[b]{0.45\textwidth}
        \includegraphics[width=\textwidth, height = 5cm, scale=1.1]{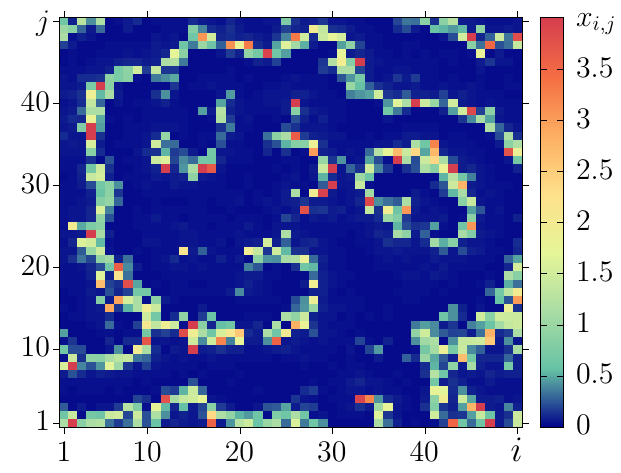}
        \caption{}
        \label{fig:spiralchimera_pattern}
    \end{subfigure}
    ~
    \begin{subfigure}[b]{0.45\textwidth}
        \includegraphics[width=\textwidth, height = 5cm,scale=1.1]{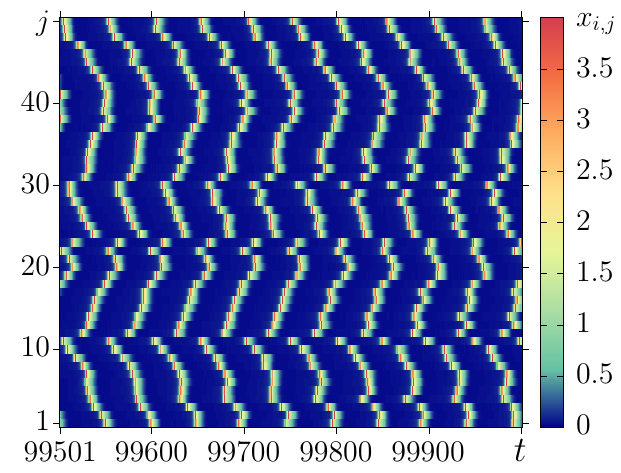}
        \caption{}
        \label{fig:spiralchimera_hm}
    \end{subfigure} 
    \caption{(a) Spiral wave chimera emerges at $\sigma$= 0.0014 and $k$ = 0.0000, The spiral wave chimera in the figure emerges within a grid structure, displaying a unique mix of synchronized and unsynchronized states radiating from a central spiral core. (b) Heatmap at $i =25$ shows the spatiotemporal evolution of the spiral wave chimera ,a coexistence of coherent and incoherent dynamics in across spatial index $j$ and time $t$, due to mechanisms of asymmetric coupling and partial synchronization inherent to chimera states.}
    \label{fig:spiralchimera}
\end{figure}

Here we showcase various types of spatiotemporal patterns observed in the 2D lattice of Chialvo neurons in the presence of electromagnetic flux and absence of delay. A target wave pattern is shown in the figure Fig. \ref{fig:targetwave_pattern}, with circular waves that are concentric and spreading outwards from the center. The main source of this type of wave is an excited media, where synchronized waves are released from the core region by oscillations with a specific phase. With the help of  flux $k$ and noise parameters $\sigma^{L}$, various dynamical behaviors, ranging from pure periodicity to total chaos, are observed on the right regime map, see Fig. \ref{fig:rm}. In the regime map, points marked in green symbols were used to indicate target waves. These waves also appear under specific conditions together with other phenomena like spiral waves and synchronized states.
\label{sec:Target Waves}
Spiral wave structure, emanates from a network lattice of Chialvo neurons where it is influenced by magnetic flux $k$ and noise parameter $\sigma^{L}$ variation in the regime map, see Fig. \ref{fig:rm}. In the latter case, patterns can occur when local interactions between components, such as magnetic fields and perturbations in the system, lead to self-organizing behaviors. The magnetic flux may drive oscillatory modulation in the system, and the variation in noise parameter $\sigma^{L}$, presumably conductivity or strength of coupling, determines the wave speed. Spiral wave is a kind of feedback mechanism such that small initial perturbations get amplified and stabilized into a rotating wave-like structure. This most likely represents oscillatory or wavefront dynamics in an evolving system. A snapshot of the spiral wave pattern is shown in Fig. \ref{fig:spiral}(a) along with a slice at a particular $i$ value passing through bthe center of the spiral wave illustrating the variaiton of the state over time.  This spiral shape points toward a non-oscillatory solution that is unstable to the strength of the magnetic fluxes, damping, or driving forces on the system under consideration.
\label{sec: Spiral Wave}
Two centered spiral wave occurs due to the interaction of local excitations in the Chialvo lattice, where two separate regions reach the threshold for wave initiation nearly simultaneously, see Fig. \ref{fig:2-center_hm}. Instead of a single excitation point dominating the system, small differences in the lattice parameters or initial conditions lead to the formation of two distinct spiral cores. The flux and sigma parameters influence the wave propagation speed and the lattice’s excitability, allowing for the coexistence of two spirals. This creates a stable pattern where waves radiate outward from both centers, each driving its own spiral wavefront across the grid. The balance of flux and noise levels at these levels prevents one center from overtaking the other, maintaining the 2-centered structure.
\label{sec: Two Centered Spiral Wave}
Multi-centered spiral wave structure forms as a result of higher excitability in the system caused by the adjusted flux $k$ and sigma $\sigma^{L}$ parameters. Multiple regions of the lattice simultaneously reach the activation threshold, each giving rise to its own spiral core, see Fig. \ref{fig:multi-center_pattern}. Instead of a single dominant spiral, the lattice becomes highly excitable, with multiple centers forming and propagating waves outward. This behavior indicates a system in a near-chaotic state, where the interactions between these spiral centers result in complex overlapping wave patterns. The increased parameter values prevent the formation of a single spiral by allowing multiple regions to act as independent excitation sources.
\label{sec: Multi-Centered Spiral Wave}
Spiral wave chimera is a dynamic phenomenon where both orderly and chaotic behavior exist side by side. In this case, we can spot areas where spiral waves are well-formed and organized, but also regions that look scattered and disorganized. This pattern occurs because the system’s excitability is finely tuned, allowing some parts to keep steady wave activity, while others fall out of sync, creating irregular motion. The specific flux and sigma values likely push the system close to chaos, leading to this intriguing blend of structured and disordered patterns, see Fig. \ref{fig:spiralchimera_pattern}. These spiral wave chimeras are interesting because they show how complex systems can simultaneously display both synchronized and random behavior.

\subsection{Influence of delay on spatiotemporal dynamics of Chialvo neurons on a 2D Lattice}
\label{sec:delay2D}
We explore the simultaneous effect of the variation of the coupling strength $\gamma$ and delay parameter $\tau$ on the transitions between various spatiotemporal patterns exhibited in the 2D lattice of Chialvo neurons.
\subsubsection{Regime map}

\begin{figure}[!h]
\hspace{-1mm}\parbox[c]{\linewidth}{ 
  \includegraphics[width=\linewidth]{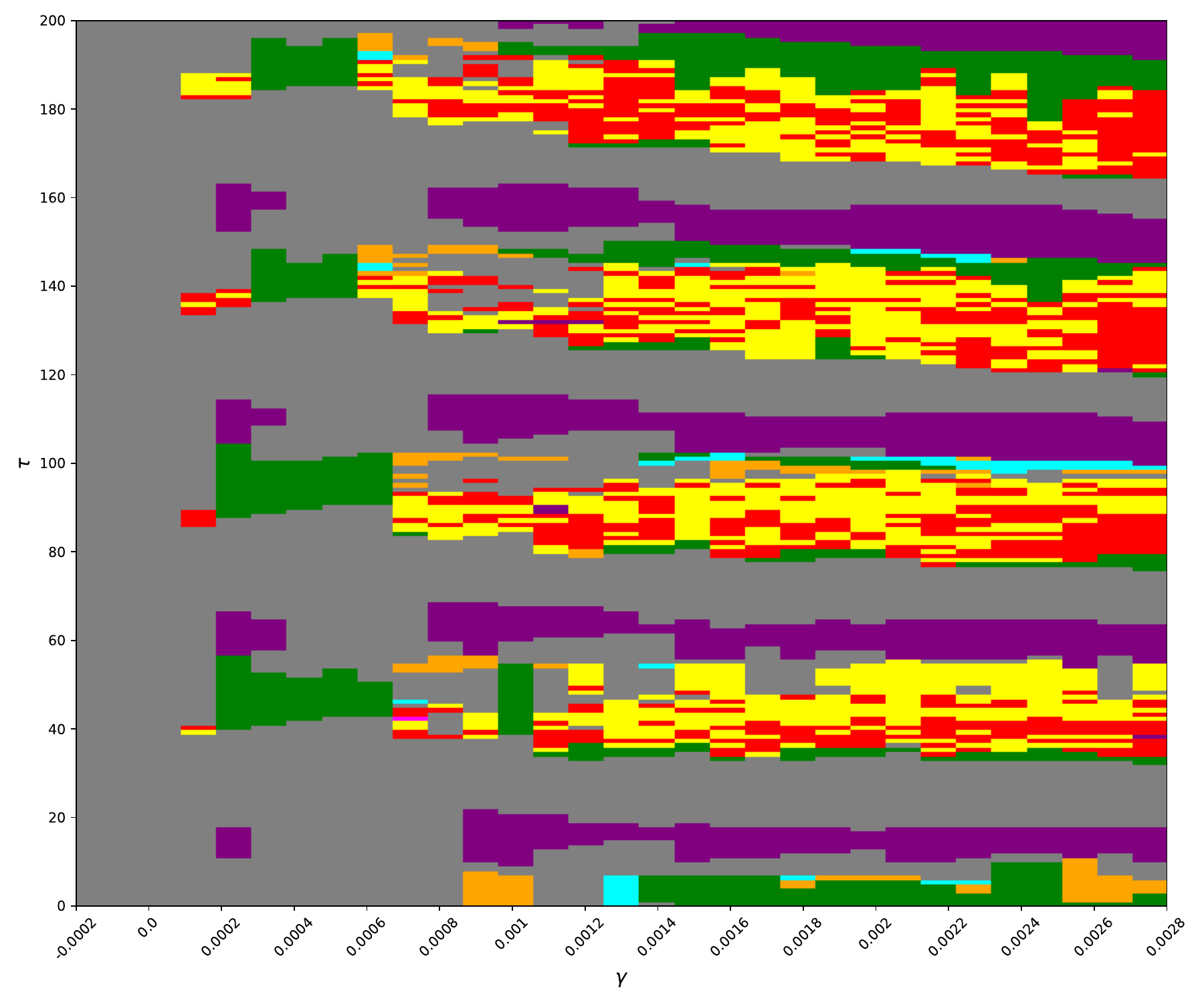}
  \vspace{-10mm} 
  \includegraphics[width=\linewidth, trim={5 mm} {25 mm} {5 mm} {40 mm}, clip]{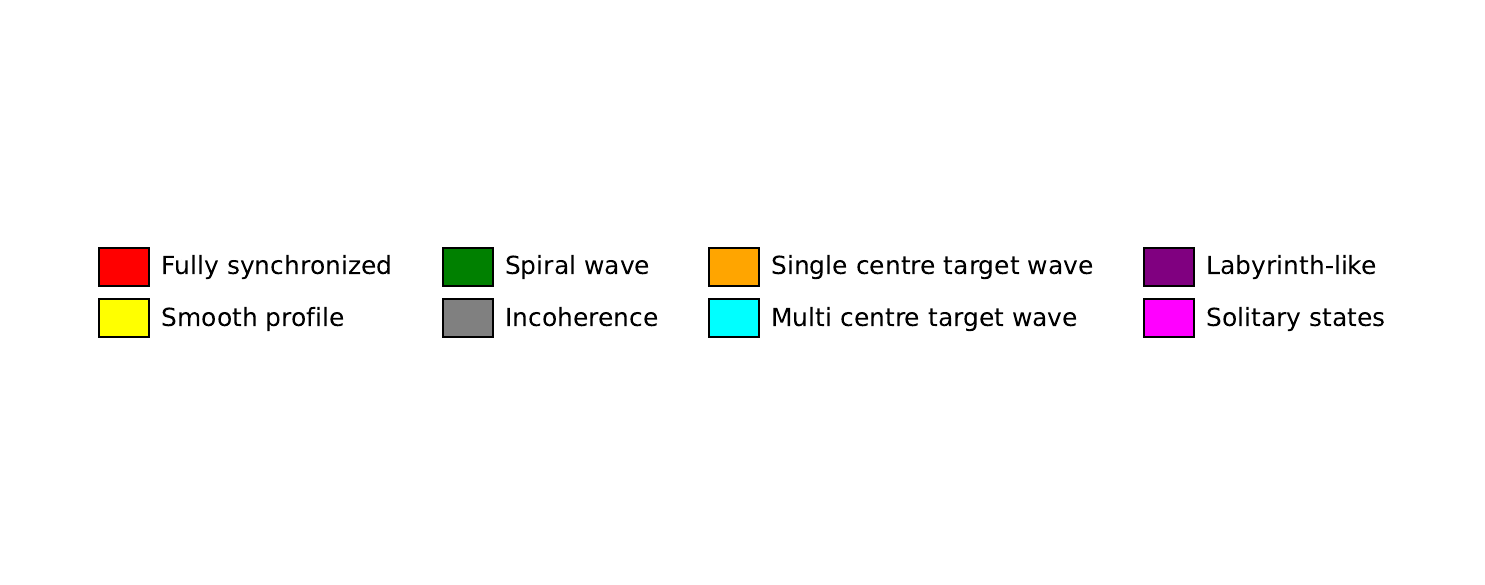}
}
\caption{\textcolor{black}{Regime map of the network Eq. (5) under the influence of delay. The plot shows the different spatiotemporal patterns observed in the 2D lattice when varying the coupling strength between -0.0002 and 0.0028 in steps of 0.0001 and the delay parameter between 0 and 200 in unit step. The red color is a fully synchronized regime; the green color corresponds to spiral waves; the yellow color is a smooth profile; the grey color is an incoherence regime; the orange color is single-center target waves; the cyan color is multi-center target waves, the magenta color corresponds to labyrinth-like structures, violet color represents solitary states. Parameters: a = 0.89 and b = 0.18, c = 0.26, P = 1, N = 50.}} 
\label{fig:regime_map}
\end{figure}

We obtain the regimes inherent to the lattice of neurons under study. The regime map is computed by simultaneously varying the values of coupling strength ($\gamma$) and delay parameter ($\tau$) within the following range: $\gamma \in [-0.0002, 0.0028]$ and $\tau \in [0, 100]$, see Fig. \ref{fig:regime_map}. We are interested in studying the effect of delay on the spatio-temporal patterns in the 2D lattice network. We show various spatiotemporal regimes that appear and disappear almost periodically as delay parameter $\tau$ is varied. This is expected in the presence of delay in a network system.  This distribution highlights the diverse range of spatiotemporal patterns that can emerge and emphasizes how crucial the presence of delay and coupling are in understanding the transformation of various spatiotemporal patterns into other regimes.

\begin{itemize}
    \item We selected a multi-centered target wave (cyan) as our initial regime without the presence of delay. We observed how spatiotemporal regimes make a transit while varying the coupling strength and delay simultaneosuly.
    \item We have observed that multi-centered target waves occur for a small increase in the magnitude of delay for the initial regime. Interestingly, it appears in the regime map (see Fig.~\ref{fig:regime_map}) when both coupling and delay are very high. 
    \item The incoherence regime (grey) covers a large area of the regime map and occurs in a wider region of the $\tau-\gamma$ parameter space.
    
    \item Prevalence of single-centered target wave (orange color) is noticed for scattered $\gamma$ values and low delay values, and their region of existence get narrower with the increase in delay.
    \item Labyrinth-like structure (magenta) is another pattern observed for moderate to high values of coupling strengths at certain intervals of delay. Note that it periodically appears and disappears along the vertical axis (delay parameter $\tau$).
    \item Similar to labyrinth-like structure, the coexistence of synchronized state (marked in red color) and smooth profile (marked in yellow) are observed for certain delay intervals with an increase in the $\gamma$ values. The above two mentioned regions appear in between the incoherence regime.
    \item Scattered spiral wave regime (green color) is also spotted in the $\gamma-\tau$ parameter plane. Interestingly, this region is observed in the close neighborhood of coexisting synchronized and smooth profiles.
\end{itemize}

\vspace{-5mm}
\subsubsection{Delay induced spatiotemporal Patterns}
In this section, we showcase diverse ranges of spatiotemporal patterns exhibited by the 2D lattice of Chialvo neurons in the presence of delay in the membrane potential variable.
\begin{figure}[h!]
\centering
\hspace{-3mm}\parbox[c]{.35\linewidth}{ 
  \includegraphics[width=\linewidth]{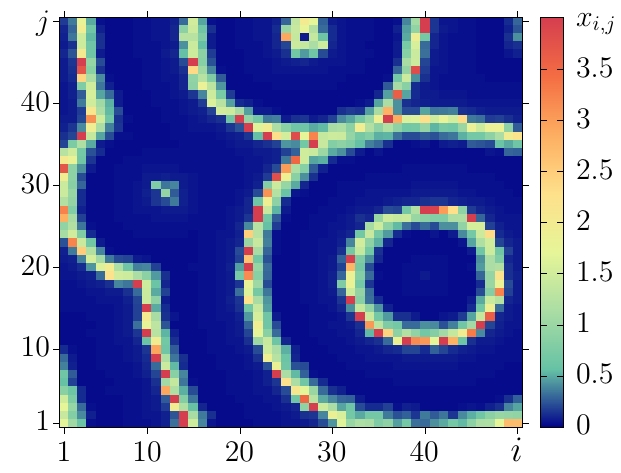}
  \vspace{-7.5mm} \center (a)
}
\hspace{1mm}\parbox[c]{.35\linewidth}{
  \includegraphics[width=\linewidth]{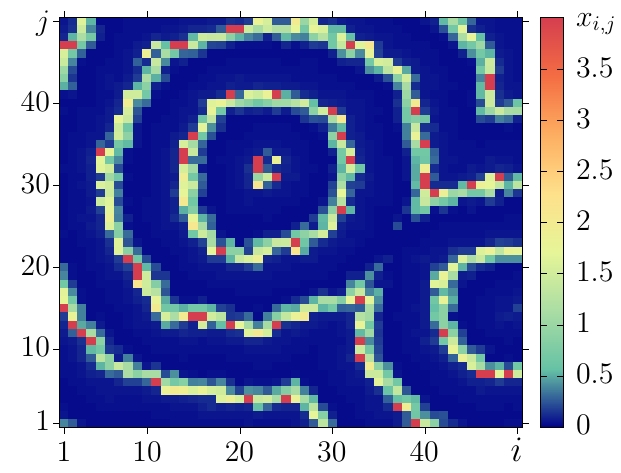}
  \vspace{-7.5mm} \center (b)
}
\caption{Snapshot of $x_{i,j}$ values for Target wave - Multi center regime with (a) $\gamma = 0.0007$ and $\tau = 46$ (b) $\gamma = 0.0013$ and $\tau = 5$ and other dynamical parameters are set as : $a=0.89, b=0.18, c=0.26, k=0.0316$.}
\label{fig:Multi-Center target wave}
\end{figure}

\begin{figure}[!h]
\centering
\hspace{-3mm}\parbox[c]{.33\linewidth}{ 
  \includegraphics[width=\linewidth]{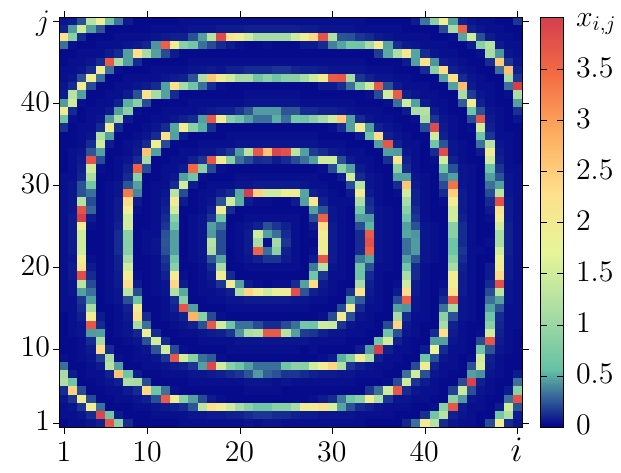}
    \vspace{-7.5mm} \center (a)
}
\hspace{1mm}\parbox[c]{.33\linewidth}{
\includegraphics[width=\linewidth]{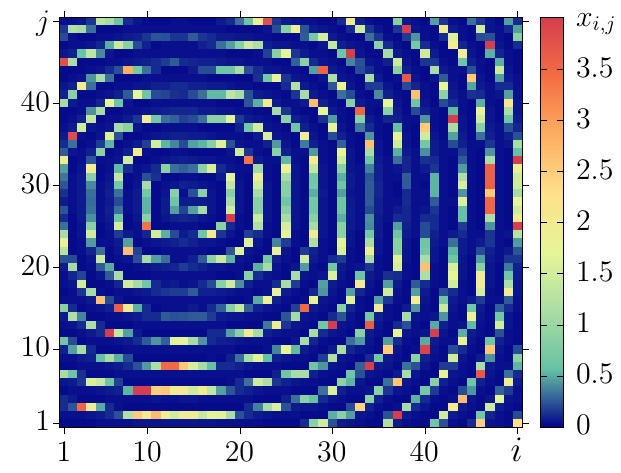}
\vspace{-7.5mm}\center (b)
}
\hspace{-4mm}\parbox[c]{.33\linewidth}{
\includegraphics[width=\linewidth]{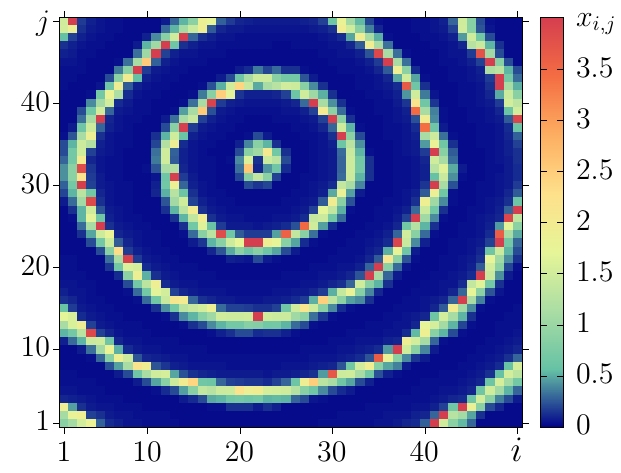}
\vspace{-7.5mm}\center (c)
}
\caption{Snapshot of $x_{i,j}$ values for Target wave regime with (a) $\gamma = 0.0018$ and $\tau = 99$ (b) $\gamma = 0.0007$ and $\tau = 7$ (c) $\gamma = 0.0012$ and $\tau = 79$ and dynamical parameters are set as: $a=0.89, b=0.18, c=0.26, k=0.0316$.}
\label{fig:Single-center Target wave}
\end{figure}

\begin{figure}[h!]
\centering
\hspace{-3mm}\parbox[c]{.35\linewidth}{ 
  \includegraphics[width=\linewidth]{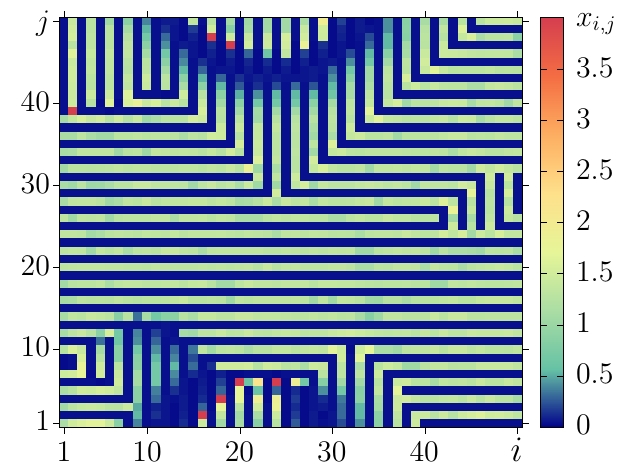}
  \vspace{-7.5mm} \center (a)
}
\hspace{1mm}\parbox[c]{.35\linewidth}{
  \includegraphics[width=\linewidth]{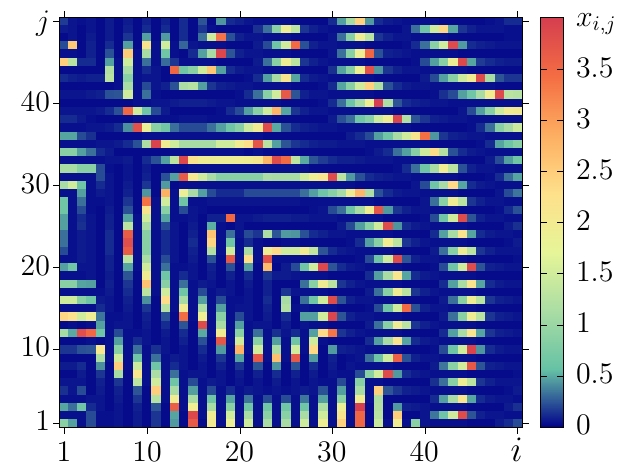}
  \vspace{-7.5mm} \center (b)
}
\caption{Snapshot of $x_{i,j}$ values for Labyrinth-like regime with a) $\gamma = 0.0009$ and $\tau = 63$ (b) $\gamma = 0.0017$ and $\tau = 63$ and other dynamical parameters set as: $a=0.89, b=0.18, c=0.26, k=0.0316$.}
\label{labyrinth}
\end{figure}

In analyzing the regime map, the target waves are likely to be found at low and high delay ranges. It was found around the presence of incoherent states, spirals and labyrinth-like structures. Multi center target waves are also observed along with single center target waves. Firstly, the multi-center target waves are observed for $\gamma$=0.0013, and later, it was found for high $\gamma$ and $\tau$ values, see Fig ~\ref{fig:Multi-Center target wave}(a-b). It displays concentric circular waves centered at the middle of the lattice, which exhibit radial synchronization and propagate outward from the center in a periodic manner.
While observing Fig ~\ref{fig:Multi-Center target wave} (b), concentric patterns are visible, with less regularity than Fig~\ref{fig:Multi-Center target wave} (a). The band's alternation indicates interference between several wave sources which is possibly due to phase mismatches introduced by the delay. In fig ~\ref{fig:Single-center Target wave}(a-c), a target wave originates from the centre, extending outward in a continuous rotational motion. The pattern indicates the coexistence of synchronization and de-synchronization
\\
From the regime map we can observe that Labyrinth-like structures (magenta) are found in between incoherence regime. These regimes are not found for the $\gamma$ values 0.0004, 0.0005, 0.0006. These regimes are most likely to occur in the delay range [9,20] and [55,68], see Fig. \ref{labyrinth}.

The patterns in the first image appear more structured and tightly packed, with fewer interruptions in the path. In the second image, labyrinth regions appear less dense, with noticeable gaps in structure, and it takes the shape of a spiral as well.
\begin{figure}[!h]
\centering
\hspace{-3mm}\parbox[c]{.35\linewidth}{ 
  \includegraphics[width=\linewidth]{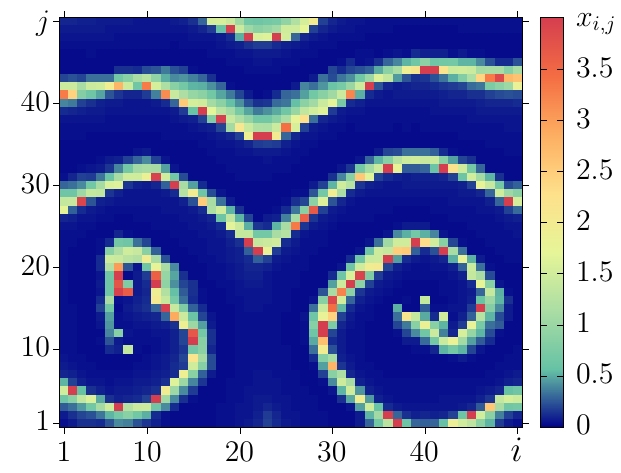}
    \vspace{-7.5mm} \center (a)
     \includegraphics[width=\linewidth]{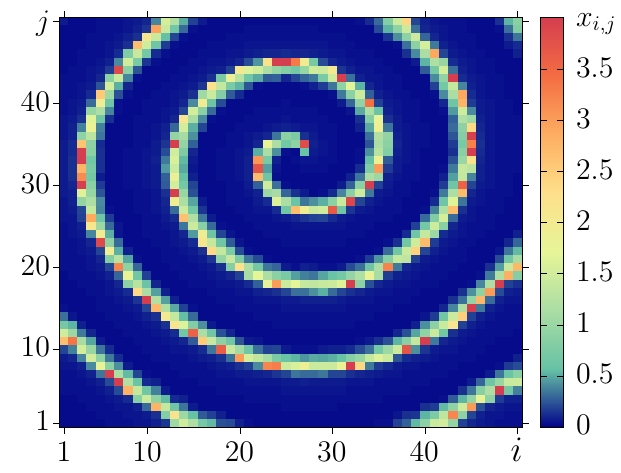}
    \vspace{-7.5mm} \center (c)
}
\hspace{1mm}\parbox[c]{.35\linewidth}{
\includegraphics[width=\linewidth]{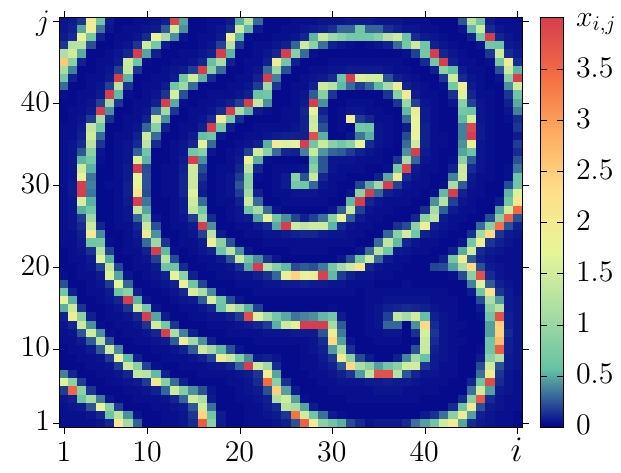}
\vspace{-7.5mm}\center (b)
   \includegraphics[width=\linewidth]{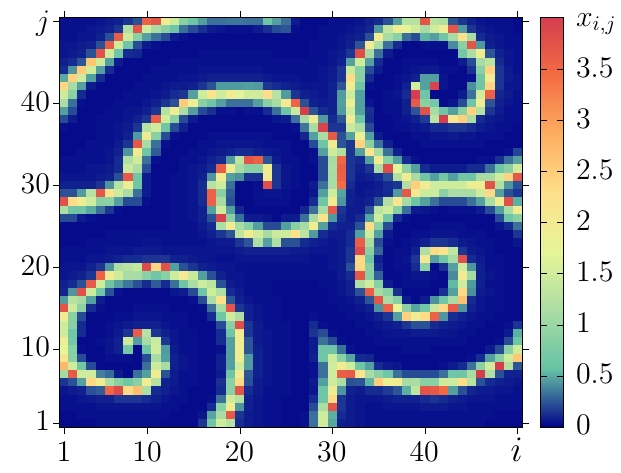}
    \vspace{-7.5mm} \center (d)
}
\caption{Snapshot of $x_{i,j}$ values for spiral waves with a) $\gamma = 0.0018$ and $\tau = 78$ (b) $\gamma = 0.0025$ and $\tau = 0$ (c) $\gamma = 0.0018$ and $\tau = 43$ (d) $\gamma = 0.0010$ and $\tau = 45$ and other dynamical parameters are set as: $a=0.89, b=0.18, c=0.26, k=0.0316$.}
\label{spiralwaves}
\end{figure}
\\
A sparse representation of the spiral wave regime is observed in the regime map. For smaller delay values, spirals are found to appear at a high value of coupling strength. Then, its prevalence in the regime map decreases as the parameters vary. Interestingly, spirals observed for larger delay values are in close proximity to synchronized and smooth profile dynamical regimes.
All the spiral waves are not stationary, meaning their centers move in the lattice space. Our analysis reveals three distinct spiral waves in the lattice of the Chialvo neuron map. The multi-center spiral wave in Fig.~\ref{spiralwaves}(a-b) is characterized by multiple initiation points, leading to complex patterns that showcase the intricate dynamics in the lattice.On the other hand, single-center spirals, depicted in Fig.~\ref{spiralwaves}(c), start from a single point and spread systematically across the lattice, reflecting a more coordinated spatiotemporal response. Multispirals with single-center are another pattern observed in the lattice. This is illustrated in Fig.~\ref{spiralwaves}(d). 


\section*{Conclusion}
In this contribution, we have explored numerically the effect of the presence of L\'{e}vy noise on the 2D lattice of Chialvo neurons interconnected in various network topologies. First, we explored various novel spatiotemporal patterns exhibited by a 2D lattice
network of Chialvo neuron 
mappings in the presence of electroamgnetic flux. It was shown that target wave pattern exists at lower values of flux values $k$ while at higher values of flux $k$, synchronized patterns were dominant. At moderate levels of noise, existence of spiral wave chimeras were observed. Noise levels beyond this worsened the dispersion of the dynamics of other spatiotemporal patterns but many multicentered spiral waves survived which indicated the complex interplay between the external electromagnetic flux interference as well as the effects of L\'{e}vy noise. 

The next network topology explored in this research is that of a ring-star network. Chialvo neuron mappings were interconnected in a hybrid ring-star network topology in the presence of L\'{e}vy noise.

Finally the effects of the presence of delay was explored in a 2D lattice of Chialvo neuron mappings.  We highlight the significant role of time delays in shaping spatiotemporal dynamics within a lattice of Chialvo neuron maps. By systematically varying the coupling strength and delay parameters, we uncovered a variety of dynamical regimes, such as synchronized states, target waves, incoherence, labyrinth-like patterns, and spiral waves. The findings stress that delays serve as a key factor, facilitating transitions between these regimes and affecting the complexity of the patterns observed. Notably, the periodic appearance of specific regimes in relation to delay highlights its potential as a tuning mechanism in excitable media.

The findings underscore the intricate interplay between coupling strength and delay in neuronal networks. Multi-center and single-center target waves and smooth profiles appear to be more sensitive to delay variations. Additionally, the prevalence of incoherent regimes at moderate coupling and elevated delays suggests that delay-induced desynchronization may have a functional role in both biological and artificial networks. Future studies could be: understanding the effect of a high magnitude of delay, $\tau > 200$. It would be interesting to account for the effect of delay on other network topologies. Leveraging machine learning approaches to classify and predict regime transitions could provide new tools for studying large-scale networks with delays. The inclusion of multilayer and multiplex networks can be carried out. This will be interesting because it is not trivial to understand the effect of L\'{e}vy noise on the interlayer and intralayer coupling strength. 
\section{Credit Author Statement}
This work was carried out by a group of students under the guidance and supervision of Dr. Sishu Shankar Muni and Dr. Andrei Bukh.
\textbf{Swetha P:}  wrote the introduction section and did the numerical simulations for 2d lattice coupled Chialvo neuron and effect of em flux (section 2.2). \textbf{J.S. Ram:} wrote the introduction and did the numerical simulations of 2d lattice coupled Chialvo neuron and the effect of delay (section 2.3). \textbf{D.S. Varghese:} reviewed and edited the introduction and did numerical simulations for section 2.3. \textbf{A.S Nair:} did the simulations for the ring network coupled Chialvo neurons (section 2.1). \textbf{A.M Suresh:} constructed the regime map for the ring network coupled Chialvo neurons (section 2.1). \textbf{C. Davis:} constructed the regime map for the ring network coupled Chialvo neurons (section 2.1). \textbf{Remya CR:} constructed the space-time diagrams of spatiotemporal patterns in the 2d lattice coupled Chialvo neurons under the influence of em flux(section 2.2). \textbf{Abhirami A.S:} constructed the regime map for 2d lattice coupled Chialvo neurons (section 2.2). \textbf{A. Hareendran:} constructed the regime map for 2d lattice coupled delay induced Chialvo neurons (section 2.3). \textbf{Fathima N:} constructed the space-time diagrams of spatiotemporal patterns in the 2d lattice coupled Chialvo neurons under the influence of em flux (section 2.2). \textbf{S.S. Muni:} supervised the project, reviewed and edited the paper,  constructed the space-time diagrams of 2d lattice coupled Chialvo neurons (section 2.2), visualized, investigated, and validated the results. \textbf{A.V. Bukh:} supervised the project, constructed regime maps of the coupled ring network of Chialvo neurons (section 2.1), reviewed and edited the paper, and validated the results.

\section*{Acknowledgements}
A. Bukh is supported by the Russian Science Foundation (Project No. 23-72-10040, \url{https://rscf.ru/en/project/23-72-10040/)}.
\section*{Conflict of Interest}
The authors confirm that this work is free from conflicts of interest.
\section*{Data Availability}
The data can be provided on reasonable request from the corresponding author.

\bibliographystyle{unsrt} 
\bibliography{Arxiv}
\end{document}